\documentclass[traditabstract]{aa} 
\usepackage{graphicx}
\usepackage{txfonts}
\newcommand{\vhbb}{$\Delta V_{\rm HB}^{\rm Bump}\,$}    
\begin{document}
   \title{The impact of an updated $^{14}N(p,\gamma)^{15}O$ reaction rate on advanced evolutionary stages of low-mass stellar models.}
\titlerunning{Revised $^{14}N(p,\gamma)^{15}O$ reaction rate and low-mass stellar models}

   \author{A. Pietrinferni\inst{1}
          \and
 S. Cassisi
          \inst{1}
          \and
          M. Salaris\inst{2}
          }

   \institute{INAF - Osservatorio Astronomico di Teramo, Via M. Maggini, 64100 Teramo, Italy\\
              \email{pietrinferni,cassisi@oa-teramo.inaf.it}
         \and
             Astrophysics Research Institute, Liverpool John Moores
University, 12 Quays House, Birkenhead, CH41 1LD, UK\\
\email{ms@astro.livjm.ac.uk}
             }

   \date{Received ; accepted}

 
  \abstract{
We have investigated the impact of the $^{14}N(p,\gamma)^{15}O$ reaction rate recently redetermined by 
the LUNA experiment, on the shell H-burning and core He-burning phases of low-mass, metal poor stellar models. 
The new reaction rate has small but noticeable effects, the largest one being a $\sim$7-8\% reduction of the red giant 
branch lifetimes. To different degrees, the lifetimes and luminosities of horizontal branch models, the mass of 
the stellar models evolving within the RR Lyrae instability strip, the luminosity of the red giant branch luminosity 
function bump, the theoretical calibrations of the R-parameter and tip of the red giant branch luminosity are also affected. 
Predictions for the tip of the red giant branch luminosity, in particular, are in 
very good agreement with the currently available empirical constraints.}


   \maketitle
%

\section{Introduction}

The last twenty years have witnessed a significant improvement in the 
calculation of the relevant input physics for stellar evolution calculations, like  
the equation of state of the stellar matter (i.e., Rogers, Swenson \& Iglesias~1996), 
radiative opacities (i.e., Iglesias \& Rogers~1996, Ferguson et al.~2005), 
nuclear cross sections (i.e., Angulo et al.~1999), 
neutrino emission rates (i.e., Haft, Raffelt, \& Weiss~1994). 
Several of these advances have been stimulated by the need to match the high-precision 
data from helioseismology, and an immediate consequence has been 
the improved accuracy of low mass-main sequence (MS) models 
for both Population I and II stars. 

A set of key ingredients in the calculation of stellar models are the nuclear reaction rates, 
and a large effort has been devoted to improve the measurements of rates at energies 
as close as possible to the Gamow peak, i.e. the energies at which nuclear 
reactions occur in stars. Thanks to these studies the reaction rates  
involved in the {\sl p-p} chain have nowadays a small uncertainty, and  
the uncertainty on the 
predicted age - luminosity relationship for low-mass MS stars  
is also negligible ($<$2\% - Chaboyer et al.~1998, Brocato et al.~1998).
However, near the end of the MS -- e.g., when the abundance of protons in the 
core is approaching zero --  
the energy supplied by the {\sl p-p} chain becomes insufficient and the 
star reacts by contracting its core to release gravitational energy. As a 
consequence, according to the virial theorem, both central temperature and density increase 
and, when the 
temperature attains a value of $\sim15\times10^6K$, the H-burning process becomes 
controlled by the {\sl CNO} cycle, whose efficiency is critically dependent 
on the $^{14}N(p,\gamma)^{15}O$ reaction rate, the slowest 
reaction of the whole cycle. 
 
Until few years ago, the rate for the $^{14}N(p,\gamma)^{15}O$ reaction was 
uncertain by a factor of 5 at least, because all available laboratory 
measurements were performed at energies well above the range of interest for 
astrophysical purposes (Angulo et al. 1999).
Recently, the LUNA experiment (Formicola et al. 2004) has significantly 
improved the low energy measurements, obtaining an estimate which is about a 
factor of 2 lower than previous determinations. 

Due to its significant impact on the efficiency of the CNO cycle, hence on 
the age - luminosity calibration during the late MS evolution, 
Imbriani et al.~(2004) studied the effect of the new reaction rate on  
the age dating of galactic 
Globular Clusters (GCs). Their analysis has shown that 
the new rate for the $^{14}N(p,\gamma)^{15}O$ reaction leads to isochrones with  
a brighter and hotter Turn Off (TO) for a fixed age, and 
the resulting age - TO luminosity calibration predicts systematically 
older GC ages, by $\sim$0.9 Gyr on average.

Weiss et al.~(2005) have investigated the effect of this new reaction rate 
on the evolution of both low- and 
intermediate-mass stars. They confirmed the results obtained 
by Imbriani et al.~(2004) and extended  
their analysis to more advanced evolutionary stages such as the He-burning 
ignition at the Red Giant Branch (RGB) tip in low-mass stars, and the core and 
shell He-burning stages in intermediate mass stars. 
More recently, Magic et al. (2010) have highlighted the crucial role played by the 
 $^{14}N(p,\gamma)^{15}O$ reaction rate in determining the transition mass 
to stars harbouring a convective core.

These studies, however, do not explore several key evolutionary properties, like the 
RGB bump brightness and evolutionary lifetimes during the 
core He-burning phase of low-mass, metal-poor stellar models, 
or the temperature location of models along the Horizontal Branch (HB). 

The main aim of this paper is to fill this gap. We analyze in detail several 
evolutionary properties of stellar models representative of stars 
currently evolving along the RGB 
and the HB of Galactic GCs. We will also investigate 
the impact of the new $^{14}N(p,\gamma)^{15}O$ reaction rate on one of the 
most important primary distance indicators, i.e. the I-band absolute   
magnitude of the tip of the RGB (TRGB), and on the R-parameter, commonly used for estimating 
the initial Helium abundance in Galactic GCs. 

The plan of this paper is as follows: in the next section we describe briefly the set of 
evolutionary models employed in our analysis; Section~3 
presents the impact of the new reaction rate 
on selected evolutionary features along the RGB stage, and provides an updated 
calibration of the TRGB brightness as standard candle; Section~4 will deal 
with the core He-burning evolutionary phase and the R-parameter calibration. 
Final remarks and conclusions close the paper.

\section{The models}

Our calculations have employed the same stellar evolutionary 
code used for the BaSTI library of stellar models (e.g., Pietrinferni et al.~2004, 
2006), the same equation of state, 
radiative opacity, neutrino energy loss and nuclear reaction rates, with the exception of 
the electron conduction opacities, that have been replaced with the recent 
results by Cassisi et al.~(2007, hereinafter C07). 
As for the $^{14}N(p,\gamma)^{15}O$ reaction, we have performed two sets of calculations,  
one employing the reaction rate used in the BaSTI code (from Angulo et al.~1999 -- hereinafter NACRE rate) and 
one with the new 
determination by Formicola et al.~(2004, see also Imbriani et al.~2005 -- hereinafter LUNA rate). 

We have computed stellar models in the mass range 
$\sim0.80M_\odot - 1.2M_\odot$, from the Zero Age Main Sequence (ZAMS) to  
the He-ignition at the RGB tip, for the  
same scaled-solar heavy element mixture adopted by Pietrinferni et 
al.~(2004 --  Grevesse \& Noels 1993), and the following choices about the 
initial chemical composition $(Y, Z)$:  (0.245, 0.0001), 
(0.245, 0.0003), (0.246, 0.0006), (0.246, 0.001), (0.248, 0.002), (0.251, 0.004), (0.256, 0.008), 
(0.259, 0.01). In the course of the paper we will use also the notation [M/H]=log($Z$/($1-Z-Y$))-
log($Z_{\odot}$/($1-Z_{\odot}-Y_{\odot}$)), where $Z_{\odot}$=0.0198 and $Y_{\odot}$=0.2734, as 
determined in Pietrinferni et al.~(2004) from the calibration of the standard solar model with the BaSTI code. 

For each chemical composition we have also computed an extended set of HB stellar models, using the standard 
method termed as \lq{reconfiguration of pre-flash models}\rq\ by Serenelli \& Weiss~(2005). In brief, 
we have considered both the He-core mass at He-ignition of a RGB progenitor whose age at the TRGB is 
of the order of 13~Gyr, and have accreted the envelope with the appropriate 
evolutionary chemical composition until the desidered total mass along the HB is attained (see, e.g.,  
Pietrinferni et al.~2004 and Serenelli \& Weiss~2005).
The RGB progenitors of these additional HB models have 
masses equal to $\sim 0.8M_\odot$ at the lowest metallicities, increasing up 
to $\sim$0.9$M_\odot$ for the more metal-rich compositions. 
We have also verified that the use of the LUNA rate for the $^{14}N(p,\gamma)^{15}O$ 
reaction does not affect the solar-based mixing length calibration, 
hence we have used the same value of the mixing length as in Pietrinferni et al.~(2004).

Finally, it is relevant to mention that the $^{14}N(p,\gamma)^{15}O$ reaction rate
has been recently reanalized by Marta et al.~(2008) taking advantage of new laboratory measurements
in the high energy regime, and an accurate analysis of the various contributions
that control the final uncertainty on the total S-factor of this reaction. 
Their recommended value for the total S-factor is however well within $1\sigma$ of the 
estimate provided by Formicola et al. (2004). Even more recently,   
Adelberger et al.~(2010) have studied 
again this reaction rate by combining together all the available measurements. They provide 
a new estimate of the total S-factor 
that is slightly larger than the LUNA result, but still in excellent agreement within the estimated uncertainty. 
Due to these results, we have also calculated selected stellar models (0.8$M_{\odot}$ 
for $Z$=0.0001 and 0.002, and 0.9$M_{\odot}$ for $Z$=0.01)   
by using alternatively the \lq{low}\rq\ and \lq{high}\rq\ rate 
for the $^{14}N(p,\gamma)^{15}O$ nuclear reaction provided by Imbriani et al.~(2006) that correspond to the $1\sigma$ 
limits on the total S-factor of the LUNA rate.

\section{The evolution along the RGB}

One of the most prominent evolutionary stages in the colour-magnitude-diagram 
of low-mass stars is the RGB, that provides also a 
valuable benchmark for low-mass 
stellar models. In fact, star counts along the RGB can be compared with predicted 
evolutionary lifetimes, the brightness of the RGB bump 
provides constraints on the extension of the convective envelope and chemical 
stratification in the model interiors, the TRGB brightness 
constrains the size of the He-core mass at the ignition of central He-burning (Renzini \& Fusi Pecci~1988).

\subsection{The RGB evolutionary lifetimes and the luminosity function bump}

During the RGB evolution the H-burning shell has a very small thickness in mass, of 
$\approx$0.001$M_\odot$ or less near the TRGB. When crossing 
the chemical discontinuity left 
over by the convective envelope, soon after the first dredge-up, the abrupt drop in molecular
weight - related to the sudden change of the H-abundance profile - affects the H-burning efficiency and,
in turn, causes a sudden change of the stellar surface luminosity.
More in detail, upon encountering the H-abundance discontinuity, 
the matter in the shell expands and 
cools slightly, causing a sharp drop in the stellar surface luminosity. 
When thermal equilibrium is restored, the stellar 
luminosity starts again to increase monotonically. As a 
consequence, the model crosses three times a narrow luminosity 
interval. This occurrence produces the characteristic bump in the RGB theoretical 
luminosity function (Thomas~1967, Iben~1968). 

Since its detection in 47 Tucanae (King el al.~1985), the RGB bump has been 
the target of several theoretical and observational investigations (see 
Di Cecco et al.~2010, and Monelli et al.~2010 for the most recent analysis, and references therein). 
The parameter widely adopted to study the RGB bump brightness is  
$\Delta V_{\rm HB}^{\rm Bump}= V_{Bump}-V_{HB}$, defined as the V-magnitude 
difference between the RGB bump and the HB at the level of the RR Lyrae 
instability strip (Fusi Pecci et al.~1990). 
From the observational 
point of view, \vhbb has the advantage to be 
independent of distance and reddening, but it is difficult to evaluate  
when the instability strip 
is scarcely populated. From a theoretical point of view, 
\vhbb depends not only on the bump level, but also on 
the exact prediction of the HB luminosity, set by the value of the He-core mass
at the He-flash (see Salaris et al.~2002, for a detailed discussion).

The first detailed comparison between predicted and observed measurements of \vhbb in 
galactic GCs was performed by Fusi Pecci et al.~(1990). By employing the theoretical 
models available at that time, these authors found that the observed dependence of 
\vhbb on the cluster metallicity was in good agreement with theory, 
but the zero point of the theoretical \vhbb values was too small by $\sim 0.4$ 
mag. For some time, this result has been considered a clear drawback of   
standard theoretical models of low-mass RGB stars. This result has been 
reanalyzed by Cassisi \& Salaris (1997), who concluded that their 
models provided a good match to the available observational data. 
More recently, this topic has been reviewed by several authors 
(e.g., Zoccali et al.~1999, Ferraro et al.~1999, Di Cecco et al.~2010), and from 
their comparisons between theory and observations one can draw the conclusion 
that lingering uncertainties on the HB theoretical brightness and the GC 
metallicity scale leave open the possibility that a discrepancy at the 
level of $\sim$0.20 mag between theory and observations may exist. 

In the following we investigate how the new rate for the 
$^{14}N(p,\gamma)^{15}O$ reaction affects the evolutionary lifetimes along 
the RGB and the brightness of the RGB bump, compared to calculations with the NACRE rate.

 \begin{figure}
\centering
\includegraphics[width=8.6cm]{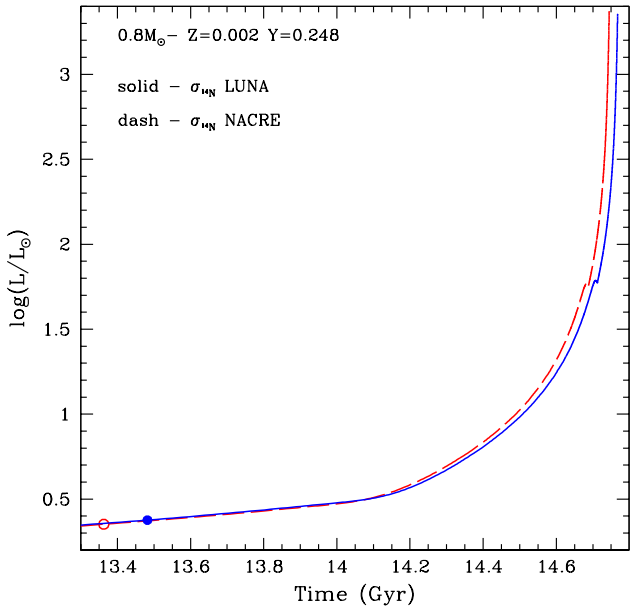}
 \caption{The evolution with time of the bolometric luminosity of selected stellar models 
(see labels), computed by using either the new LUNA $^{14}N(p,\gamma)^{15}O$ reaction rate, or the NACRE one. 
The full and open dots mark the end of the central H-burning for the the model computed with the LUNA and NACRE  
rate, respectively.}
\label{ltime}
\end{figure}

Figure~\ref{ltime} displays the bolometric luminosity as a function 
of time from the end of the core H-burning stage to the TRGB, for 
a post-MS stellar model of a typical GC star (M=0.8$M_{\odot}$, $Y$=0.248, $Z$=0.002).   
As expected from previous 
investigations (Imbriani et al.~2004, Weiss et al.~2005) the model 
computed with the new reaction rate reaches the end of the H-burning stage 
with an age larger by about 120~Myr ($\sim$1\% age difference) compared to results with 
the NACRE rate. 
 \begin{figure}
\centering
\includegraphics[width=8.6cm]{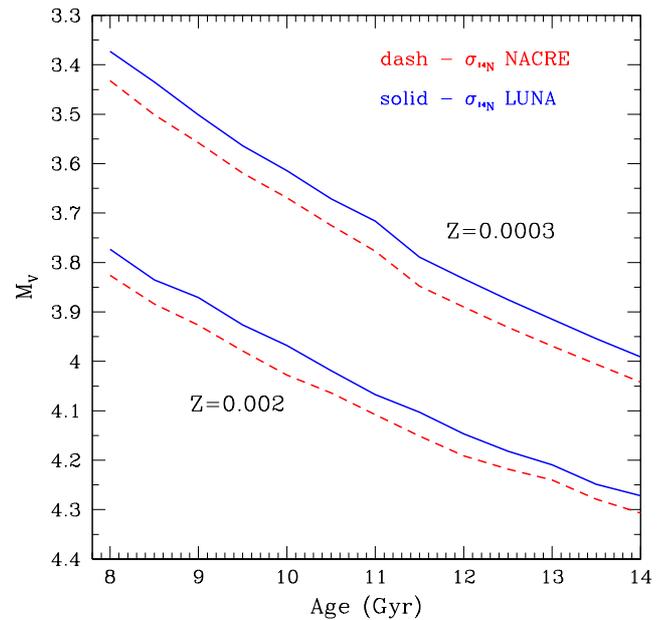}
 \caption{The absolute V-band magnitude of the TO as a function of age, for theoretical isochrones 
with Z=0.0003 and Z=0.002, as derived from stellar models computed with either the LUNA or the NACRE 
$^{14}N(p,\gamma)^{15}O$ reaction rate. }
\label{mvage}
\end{figure}
The impact of this age difference on the $M_V^{TO}$-age relation in GC isochrones can be estimated 
from Fig.~\ref{mvage}. The $M_V^{TO}$-age relation obtained using the LUNA rate is systematically 
brighter compared to results with the NACRE rate. As a consequence, for a fixed observed value of 
$M_V^{TO}$, LUNA isochrones give ages higher by $\sim$0.7~$Gyr$, compared to calculations with the NACRE rate.

When the models displayed in Fig.\ref{ltime} reach the He-flash,  
the age difference between calculations with the NACRE or the LUNA rate 
is reduced from $\sim$120~Myr to $\sim$20~Myr, implying a slightly faster RGB evolution when 
the LUNA reaction rate is employed. 

 \begin{figure}
\centering
\includegraphics[width=8.6cm]{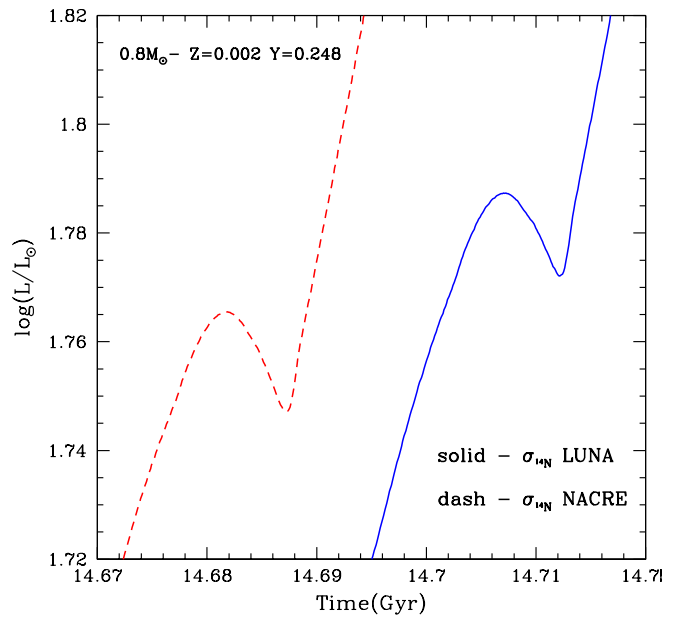}
 \caption{As in Fig.~\ref{ltime}, but centred around the region of the RGB bump.}
\label{bump1}
\end{figure}

\begin{figure}
\centering
\includegraphics[width=8.6cm]{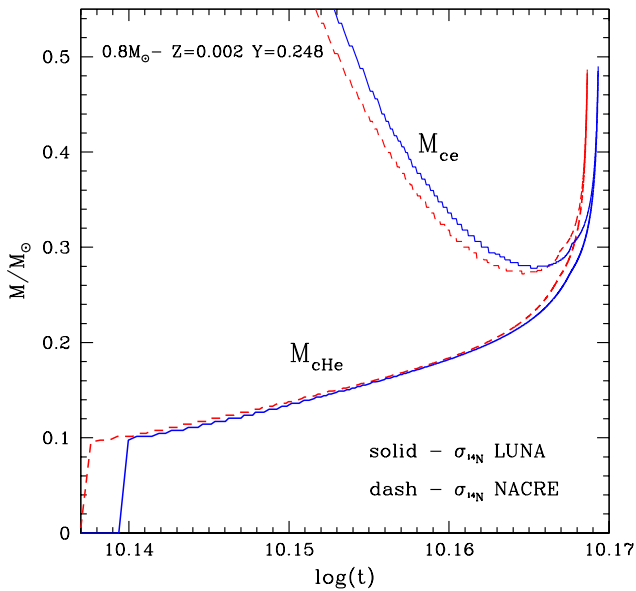}
 \caption{The time evolution of the He-core mass and the location of the bottom of the 
 convective envelope, for the same models shown in Fig.~\ref{bump1}.}
\label{bump2}
\end{figure}

Figure~\ref{bump1} shows the evolution of the luminosity with time in   
the region of the RGB bump. The LUNA rate causes a brighter bump 
(by $\sim 0.06$ mag), due to a slightly shallower convective envelope 
at the first dredge up, as shown by Fig.~\ref{bump2}.
This is due to the the fact that the LUNA based model has 
an hotter H-burning shell for a fixed He core mass.

\begin{figure}
\centering
\includegraphics[width=8.6cm]{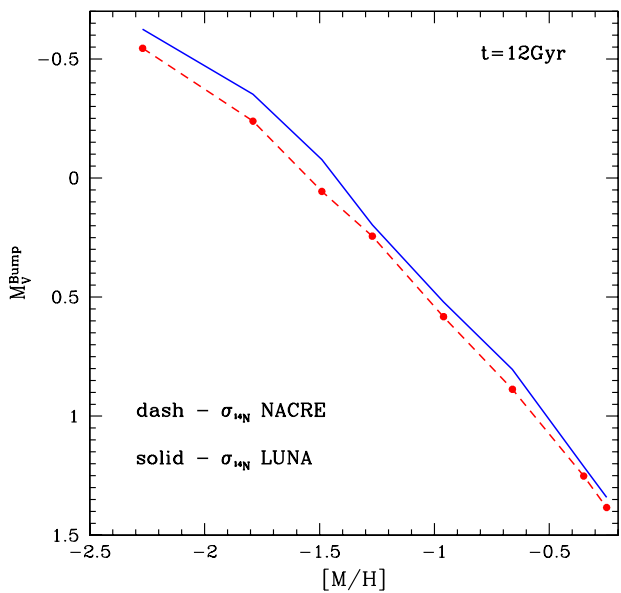}
 \caption{The V-band magnitude of the RGB bump as a function of the metallicity, 
for a stellar population with age equal to 12~Gyr, as derived from stellar
  models computed with either the LUNA or the NACRE  
  $^{14}N(p,\gamma)^{15}O$ reaction rate. }
\label{bump3}
\end{figure}

Finally, Fig.~\ref{bump3} shows the trend of the absolute visual magnitude of the 
RGB bump as a function of the metallicity for a 12~Gyr old stellar population. 
The LUNA rate causes a systematically brighter RGB bump in the models, for the 
whole Galactic GC metallicity range. The difference with respect 
to results with the NACRE rate is of the order of 0.07 mag at the lowest metallicities, becoming 
almost negligible in the metal rich regime.


\subsection{The RGB tip}

The evolution along the RGB ends when the thermal conditions 
for igniting helium burning are attained in the electron degenerate core.
A runaway nuclear burning of He in 
the core occurs, the so-called He-flash, after which the star contracts and 
starts quiescent core helium burning along the HB. 
The value of the He-core mass at the He-flash ($M_{cHe}$) 
controls the brightness of both the 
TRGB and the HB, two of the most important primary standard candles for 
old stellar populations (see, e.g., Lee, Freedman \& Madore~1993, Salaris \& Cassisi~1997, 
Caputo~1998, Catelan~2009, and references therein).
Table~\ref{tab1} lists some relevant evolutionary properties at the TRGB 
for selected models computed with either the LUNA  
or the NACRE rate. The LUNA reaction rate causes  
an increase of the He-core mass at the He-flash of the order of 
$\sim0.002-0.003M_\odot$, compared to the results with the NACRE rate. 
The size of this difference is only slightly larger than the change due to the use of different 
sets of conductive opacities (Salaris et al.~2002, Cassisi et al.~2007). 
 
Although the He-core mass is larger in the LUNA models, the surface luminosity 
at the TRGB is lower, the difference being $\Delta\log(L/L_\odot)\approx 0.02$~dex. 
This is due to the lower efficiency of the CNO cycle in 
the H-burning shell, that offsets the luminosity increase expected from the higher 
core mass (see, e.g., the discussion in Weiss et al.~2005).

Weiss et al.~(2005) obtained qualitatively similar results in terms of variation of 
$M_{cHe}$ and luminosity at the TRGB when updating 
their $^{14}N(p,\gamma)^{15}O$ reaction rate to the LUNA value and   
using the NACRE 3$\alpha$ reaction rate. Quantitatively, however, both  
the increase of $M_{cHe}$ and decrease of the RGB tip luminosity are 
(at least for the Z=0.0001 case tabulated in their Table~1) 
about factor of 2 larger. This may be due to the fact their reference 
$^{14}N(p,\gamma)^{15}O$ reaction rate was the Adelberger et al.~(1998) one, that is 
 even larger than our reference NACRE rate.

Before closing this section, we comment briefly on the results obtained 
with the \lq{low}\rq\ and the \lq{high}\rq\ estimates of the LUNA rate (see Sect.~2). 
The TRGB bolometric 
luminosity is affected at the level of $\Delta\log(L/L_\odot)\approx \pm0.003$, while the He core mass
at He-ignition is modified by only $\pm0.0004M_\odot$. As for the evolutionary timescales, both 
MS and RGB lifetimes are affected at the level of at most 1\%.
These results show how the uncertainties on the recent estimates of the $^{14}N(p,\gamma)^{15}O$ reaction rate 
have only a very small impact on the properties of low-mass stellar models.

\begin{table}
\caption{Selected properties of TRGB models, computed by 
adopting either the LUNA reaction rate, or the NACRE one.}            
\label{tab1}     
\centering                         
\begin{tabular}{l c c}      
\hline\hline                 
  &  NACRE & LUNA \\    
\hline 
\smallskip
 & \multicolumn{2}{c}{$0.8M_\odot - Z=0.0001 - Y=0.245$}\\   
\hline       
    $\log(L/L_\odot)$  & 3.261   & 3.245   \\
    $\log(T_{eff})$   & 3.643   & 3.644  \\
    Age(Myr)  & 12344,36   & 12361.32   \\
     $M_{cHe}/M_\odot$     &  0.4968    &  0.5000   \\
     $Y_{env}$ & 0.253    &  0.252 \\
     \hline  
\smallskip
 & \multicolumn{2}{c}{$0.8M_\odot - Z=0.002 - Y=0.248$}\\   
\hline       
         $\log(L/L_\odot)$ &    3.367   & 3.352 \\ 
         $\log(T_{eff})$     &    3.574   & 3.574 \\
          Age(Myr)            & 14744.69      & 14767.70 \\  
          $M_{cHe}/M_\odot$  &  0.4825  &  0.4852  \\ 
          $Y_{env}$                 &  0.263    &  0.262    \\
\hline 
\smallskip
 & \multicolumn{2}{c}{$0.9M_\odot - Z=0.01 - Y=0.259$}\\   
\hline       
          $\log(L/L_\odot)$  &  3.413   & 3.399\\
          $\log(T_{eff})$        & 3.516   & 3.518 \\
          Age(Myr)              & 14529.97   & 14539.97\\   
   $M_{cHe}/M_\odot$    &  0.4737  &  0.4760 \\
          $Y_{env}$             &    0.279  &  0.279 \\
\hline    
\end{tabular}
\label{tab1}
\end{table}

\subsubsection{The calibration of the TRGB method}

The I-band TRGB magnitude is one of the most powerful primary 
distance indicators. For a careful discussion of the observational advantages 
and drawbacks of this standard candle we refer to Lee et al.~(1993) and 
Madore \& Freedman (1995), while an accurate analysis of the uncertainties 
affecting the theoretical calibration of this distance method can be found 
in Salaris et al.~(2002) and Cassisi~(2005).

On the observational side, Bellazzini et al.~(2001) have recently published an 
empirical calibration based on the GC $\omega$~Cen. They adopted the 
distance obtained by Thompson et al.~(2001) from the analysis of one eclipsing 
binary system belonging to the cluster, and determined the TRGB apparent I-band brightness 
from a well populated CMD, deriving an absolute magnitude $M_I^{TRGB}=-4.04\pm0.12$ 
in the I-Cousins filter. To date, 
all theoretical calibration of the TRGB method are within $1.5\sigma$ of 
this calibration as shown by Cassisi~(2010). More recently, the empirical 
calibration of the TRGB method has been extended to larger metallicities and 
also to the near infrared bands by Bellazzini et al.~(2004), with the inclusion of new 
observations for the massive GC 47~Tuc. Moreover, Bellazzini et al.
~(2004) repeated the analysis of $\omega$~Cen with the new  
study of its extinction by Lub~(2002), and derived 
an almost identical value, $M_I^{TRGB}=-4.05\pm0.12$.

 \begin{figure}
\centering
\includegraphics[width=8.6cm]{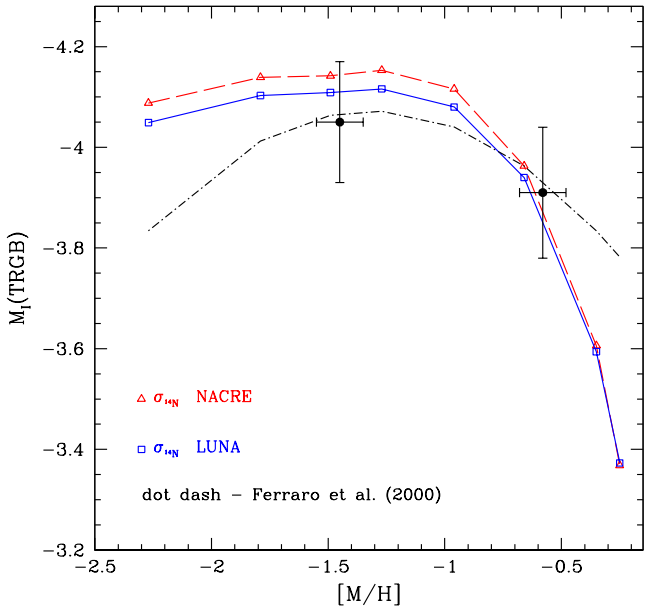}
 \caption{Several calibrations of the TRGB magnitude in the I-Cousins band,  
  as a function of the  metallicity. We display results based on the LUNA and NACRE 
 $^{14}N(p,\gamma)^{15}O$ reaction rate, and the semi-empirical calibration by Ferraro et 
 al.~(2000). The two filled circles with 
 error bars correspond to the empirical estimates in 
 the $\omega$~Cen and 47~Tuc, obtained by Bellazzini et al.~(2004).}
\label{tip1}
\end{figure}

Figure~\ref{tip1} shows three calibrations of the 
TRGB magnitude in the I-Cousins band as a function of the metallicity, 
compared with Bellazzini et al.~(2004) empirical estimates\footnote{Bolometric luminosities have been 
transformed to broadband magnitudes using the bolometric corrections described in Pietrinferni et al.~(2004).}

When using the LUNA rate instead of the NACRE one, 
$M_I^{TRGB}$ becomes fainter by about 0.05~mag at [M/H]$\leq -$1.0, 
while at higher metallicites the differences are smaller. Notice that 
the calibration with the LUNA rate is 
within $\sim 0.5\sigma$ of the empirical estimates for both $\omega$~Cen and 47~Tuc.

\begin{figure}
\centering
\includegraphics[width=8.6cm]{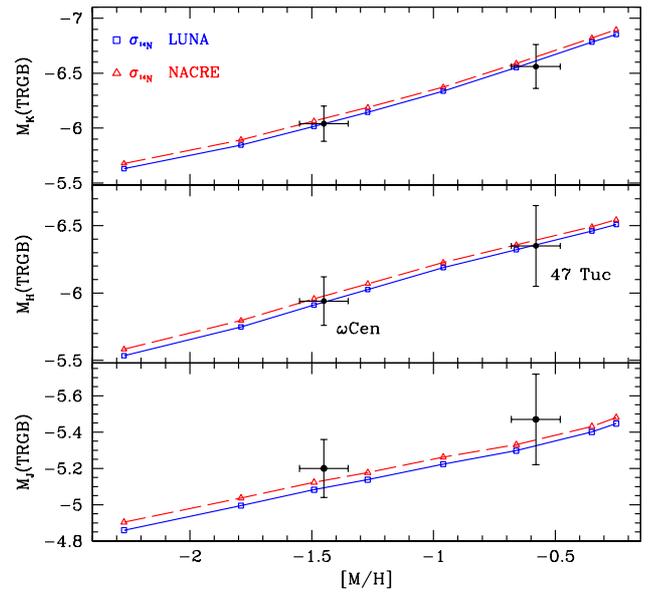}
 \caption{As in Fig.~\ref{tip1} but for the JHK near-infrared bands.  }
\label{tip2}
\end{figure}

Figure~\ref{tip2} displays a similar comparison but in the near infrared. Models 
calculated with the LUNA rate are generally fainter by $\sim$0.05~mag, and 
agree well with empirical data also 
in the H- and K-bands, while they are more discrepant -- but still lie within the error bars -- 
in the J-band. This may suggest a source of uncertainty in the 
bolometric correction to the J-band (see also the discussion in Bellazzini~2007).

The best fits of our theoretical calibrations with the LUNA reaction rate are the following:

\begin{equation}
M_I= -2.912 + 2.375\cdot[M/H] + 1.480\cdot[M/H]^2 + 0.289\cdot[M/H]^3
\end{equation}

\begin{equation}
M_K= -7.063 - 0.844\cdot[M/H] - 0.093\cdot[M/H]^2
\end{equation}

\begin{equation}
M_H= -6.638 - 0.487\cdot[M/H]
\end{equation}

\begin{equation}
M_J= -5.501 - 0.284\cdot[M/H]
\end{equation}

 \begin{figure}
\centering
\includegraphics[width=8.6cm]{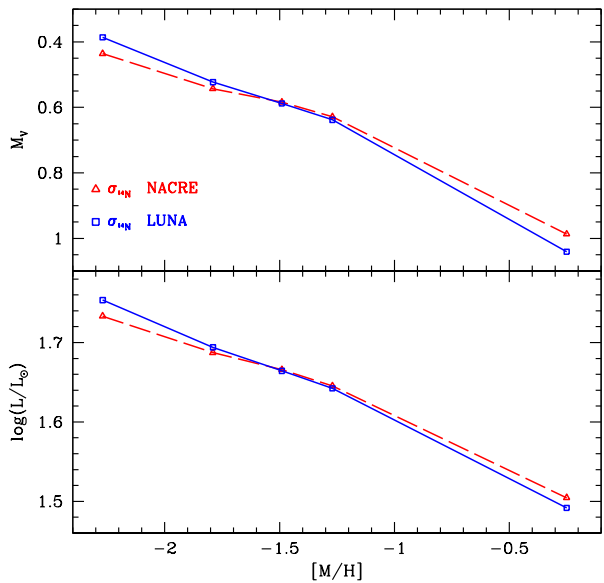}
 \caption{{\it Upper panel}: Absolute visual magnitude of the ZAHB at 
 $\log{T_{eff}}=3.85$, as a function of the metallicity. The stellar 
 models have been computed by adopting alternatively the NACRE or the LUNA  
 $^{14}N(p,\gamma)^{15}O$ reaction rate. {\it Lower panel}: 
 As in the upper panel, but for the bolometric luminosity of the ZAHB.}
\label{zahb}
\end{figure}

\section{The core He-burning stage}

As discussed in the previous section, the use of the new LUNA reaction rate 
affects the mass of the He core at the RGB Tip. One can therefore expect 
that the brightness of the Zero Age Horizontal Branch (ZAHB) is also affected.
Figure~\ref{zahb} compares the predicted bolometric luminosity and 
V-band absolute magnitude of ZAHB models at 
$\log({T_{eff}})=3.85$ - a point taken as representative of the average 
effective temperature of the RR Lyrae instability strip - as a function of 
metallicity, for various assumptions about the rate of the 
$^{14}N(p,\gamma)^{15}O$ nuclear reaction. 
The updated LUNA rate does not affect the brightness of the ZAHB 
at intermediate metallicities, while it has a modest impact,  
at the level of $\sim 0.05$~mag, in 
the high- and low-metallicity regime: models based on the 
LUNA rate appears brighter at the lowest metallicity and fainter at the opposite 
end of the metallicity spectrum. 
Given that the ZAHB luminosity increases with increasing $M_{cHe}$ and, at fixed $M_{cHe}$, 
with increasing the efficiency of the H-shell burning, 
this behaviour can be explained as follows. At low metallicites, 
because of the low contribution of the CNO burning to the surface luminosity in 
models within the instability strip, 
it is the increase of $M_{cHe}$ that causes the difference between LUNA and NACRE models.
Increasing the initial metallicity enhances the contribution of the H-burning shell 
to the output luminosity, and in this case it is the 
decreased efficiency of the CNO burning with the LUNA rate that progressively dominates 
the behaviour of the ZAHB luminosity, compared to NACRE results.

As a consequence of the ZAHB and RGB calculations with LUNA and NACRE rates, 
one can conclude that the use of the LUNA rate  
leaves unaffected the theoretical calibration of the \vhbb parameter 
for [M/H] below $\sim-$1.8, decreases \vhbb by $\sim$0.05~mag at intermediate metallicites, 
and up to $\sim$0.1~mag when [M/H] increases above $\sim-$1.

 \begin{figure}
\centering
\includegraphics[width=8.6cm]{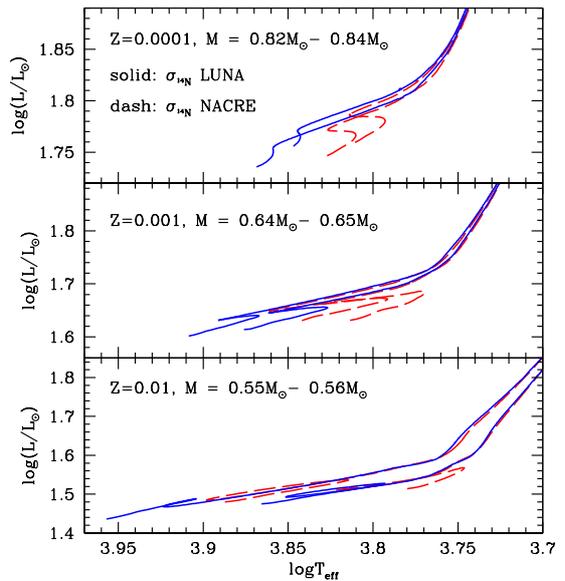}
 \caption{H-R diagram of selected HB evolutionary tracks, computed by 
 using the LUNA reaction rate (solid line) or the NACRE one (dashed line), and for various 
 assumptions about the metallicity and total mass.}
\label{hb}
\end{figure}

Regarding the $T_{eff}$ location and morphology of HB evolutionary tracks, 
Fig.~\ref{hb} shows the H-R diagram of 
selected models computed with the NACRE and LUNA rates, 
for three different metallicities. 
Given that models computed with the LUNA rate 
have a larger $M_{cHe}$ at the He-flash, they will also have a 
hotter location in H-R diagram for a given value of the total stellar mass. 
As a consequence, the mean mass of models within the RR Lyrae  
instability strip is increased when the LUNA rate is adopted, by an amount ranging from ~0.05$M_\odot$ 
for Z=0.001, to ~0.007$M_\odot$ for Z=0.01.

Finally, we find that the effect of the LUNA rate on the core He-burning 
lifetime is small, amounting to a reduction of 
$\sim3-4$\% with respect to models based on the NACRE rate. 
This difference has been evaluated for models whose ZAHB location 
is within the instability strip. For lower mass HB models  
the variation is even smaller ($\sim1$ Myr).

\subsection{The R-parameter calibration}
\begin{table}
\caption{Values of the R-parameter as obtained for various assumptions about the 
$^{14}N(p,\gamma)^{15}O$ nuclear reaction rate and a reference age of 13~Gyr. 
The values obtained by the BaSTI library are also listed.}            
\label{tab2}
\centering                         
\begin{tabular}{c c c c c}      
\hline\hline                 
  Z & Y & $R_{NACRE}$ & $R_{LUNA}$ & $R_{BaSTI}$\\    
\hline 
  0.0001 &  0.245  &  1.411 & 1.448  &  1.433\\
  0.0003 &  0.245  &  1.420 & 1.410  &  1.422\\
  0.0010 &  0.246  &  1.427 & 1.393  &  1.404\\
  0.0100 &  0.259  &  1.885 & 1.730  &  1.856\\
\hline    
\end{tabular}
\label{tab2}
\end{table}

A powerful method to estimate the initial He-content 
in GC stars is based on the so-called R-parameter (Iben~1968, 
Salaris et al.~2004 and references therein). This parameter is defined as the 
ratio of the star counts along the HB to the number of RGB stars brighter than 
the HB. Theoretically, the value of the R-parameter 
can be obtained from the ratio between the HB lifetime of a model  
whose ZAHB location is within the RR Lyrae instability strip,  
and the time spent by its RGB progenitor 
at magnitudes brighter than the ZAHB brightness: 
$R=t_{cHe}/t_{RGB}$. Given 
that the lifetime during the HB phase depends on the total stellar 
mass, hence on the ZAHB effective temperature (see Castellani et al.~1994), 
it is common in case of GCs with very blue HB morphologies, to apply 
a correction that takes into account the increase of $t_{cHe}$ with decreasing HB mass 
(see Cassisi et al.~2003 and Sandquist et al.~2010).

As discussed before, the use of the new LUNA reaction rate affects - albeit slightly - 
ZAHB and TRGB brightness, HB and RGB evolutionary lifetimes, and it is important to 
verify whether the theoretical calibration of the R-parameter is influenced by the 
use of this updated rate. To this purpose, we have 
compared the calibrations of the R-parameter as a function of $Z$ and $Y$ 
obtained with the LUNA and NACRE rates, keeping all other inputs unchanged. 
For some selected metallicities, the values of the R-parameter are listed in Table~\ref{tab2}, where for the sake
of comparison also the values based on the BaSTI models are provided.
Given that dR/d$Y \sim$10, we find that, for a given value of R, the LUNA calibration provides 
helium mass fractions $Y$ lower by 0.004 at $Z=0.0001$, but are higher by 0.001, 0.003 and 
0.015 at $Z= 0.0003$, 0.001, and 0.01, respectively.

Overall, at fixed $Z$, the change of R induced by the use of the new
LUNA reaction rate is smaller than typical error bars on individual empirical estimates of R, 
and smaller than the theoretical uncertainty coming from the $^{12}C(\alpha,\gamma)^{16}O$ reaction rate 
(Cassisi et al.~2003). 

It is interesting to check how these variations would impact on the estimate of the primordial He abundance
based on measurements of the R-parameter in metal-poor galactic GCs. Salaris et al.~(2004) --  
expanding upon the work by Cassisi et al.~(2003) -- determined the initial He mass fraction ($Y_{GGC}$) of stars in a sample 
of 57 Galactic GCs (assuming stars within the same cluster share the same initial He-content).  
They derived an average $Y_{GGC}=0.250\pm0.006$, and no significant trend with metallicity. 
Selected values of their theoretical calibration of R are 
displayed in Table~\ref{tab2}. The model input physics is the same as in the BaSTI library, i.e. differs 
from the NACRE values in the table because of the different electron conduction opacities used. The BaSTI models 
have been calculated using the Pothekin~(1999) electron conduction opacities, instead of the C07 ones employed here. 

The LUNA calibration displayed in Table~\ref{tab2} gives $Y_{GGC}\sim0.251$ at the lowest and intermediate 
metallicities of Salaris et al.~(2004) GC sample. This value is consistent with the primordial He-content 
obtained from standard Big Bang nucleosynthesis calculations (see Steigman~2010). 
At the high metallicity end of the GC sample 
([Fe/H] between $-$0.3 and $-$0.6, the exact value depending on the adopted metallicity scale) the estimated 
initial He mass fraction increases to $Y_{GGC}\sim0.261$.

\section{Summary}

We have investigated the impact of the updated $^{14}N(p,\gamma)^{15}O$ reaction rate from the LUNA experiment 
on post-MS stages of low-mass stellar models. In agreement with previous works on the same subject, we have 
found that the new reaction rate has overall a small influence on the models, although some evolutionary properties 
are more sizably affected.
Our results can be summarized as follows:

\begin{itemize}

\item the use of the updated LUNA reaction rate causes a slower MS evolution but a 
7-8\% shorter lifetime along the RGB, compared 
to results with the NACRE rate. 
As a consequence, for a fixed value of the mass and initial chemical composition, 
LUNA low-mass stellar models attain the He-ignition at the TRGB 
with approximately the same age of models based on the NACRE rate;

\item for a fixed age and initial chemical composition, 
the bump in the RGB luminosity function predicted by theoretical isochrones  
is brighter by $\sim0.06$~mag with the new rate, 
due to slightly shallower convective envelopes at the $1^{th}$ dredge up;

\item the bolometric luminosity of the TRGB for old stellar populations 
is lower than in models based on the NACRE rate, despite the slightly larger He-core mass at the TRGB. 
The new LUNA rate provides a new theoretical calibration of the I-band 
TRGB magnitude as a function of [M/H], that is in good agreement with robust empirical constraints. The same result holds for  
near-infrared bands; only the J-band theoretical calibration seems to be slightly discrepant, 
although always within $1\sigma$ of the empirical data. 
We provide analytical calibrations of the $I, J, H$ and $K$ TRGB absolute magnitudes as a function of [M/H];

\item the new LUNA rate affects 
also the ZAHB brightness, the size of the effect depending on the metallicity: at the lowest metallicities 
ZAHB models are brighter by about 0.05~mag, while they appear fainter by the same amount 
at high metallicities, leaving unaffected the intermediate regime. 
When combining this result with the effect on the RGB bump brightness, one obtains that the new theoretical 
calibration of the \vhbb parameter is unaffected in the low metallicity regime, while it is decreased by $\sim 0.05$ mag 
and $\sim 0.10$ mag in the intermediate- and
metal-rich regime, respectively;

\item the core He-burning lifetimes for models within the RR Lyrae instability strip are 
decreased by $\sim3-4$\%. The same variation is found for models located at the red side of the strip, whereas 
the effect is smaller for models blueward of the strip;

\item we have investigated the effect of the new rate on the ca\-li\-bra\-tion of R-parameter, a powerful indicator for the initial He-abundance 
in old stellar populations. R-parameter values based on models relying on the LUNA rate provide, for a fixed value of R, 
He mass fractions lower by about 0.004 at $Z$=0.0001, but higher by 0.001, 0.003 and 0.015 
at $Z$=0.0003, 0.001 and 0.01, respectively. 

\end{itemize}

\begin{acknowledgements}

We warmly thank our referee (Achim Weiss) 
for very helpful comments that have helped improve the presentation of our results.
This research has made use of NASA's
Astrophysics Data System Bibliographic Services, which is operated by the Jet
Propulsion Laboratory, California Institute of Technology, under contract with the
National Aeronautics and Space Administration. A.P. and S.C. acknowledge the financial
support of the Ministero della Ricerca Scientifica e dell'Universita' PRIN MIUR
2007: \lq{Multiple stellar populations in globular clusters}\rq\ (PI: G. Piotto),
ASI grant ASI-INAF I/016/07/0, and the Italian Theoretical Virtual Observatory
Project (PI: F. Pasian).
\end{acknowledgements}

\end{document}